\documentclass[]{raa}
\usepackage{graphicx,times}
\usepackage{natbib}

\newcommand{\dft}{$T_{\rm df}$~}
\newcommand{\Rom}{\rm R_m}
\newcommand{\Rin}{\rm R_m^{-1}}
\newcommand{\lld}{$\ln \Lambda$~}
\newcommand{\CL}{Coulomb logarithm~}

\begin{document}

   \title{Modelling the dynamical friction timescale of sinking satellite 
          $^*$
      \footnotetext{\small $*$ Supported by the National Natural
              Science Foundation of China.}
         }

 \volnopage{ {\bf 2010} Vol.\ {\bf X} No. {\bf XX}, 000--000}
   \setcounter{page}{1}

   \author{Jianling Gan
      \inst{1,3}
   \and Xi Kang
      \inst{2}
   \and Jinliang Hou
      \inst{1}
   \and Ruixiang Chang
      \inst{1}
   }

   \institute{Key Laboratory for Research in Galaxies and Cosmology,
    Shanghai Astronomical Observatory,\\
    Chinese Academy of Sciences, 80 Nandan RD, Shanghai,  200030, China\\
    {\it jlgan@shao.ac.cn} \\
        \and
        The Purple Mountain Observatory, 2 West Beijing Road,
    Nanjing 210008, China\\
        \and
        Graduate School of the Chinese Academy of Sciences,
        No.19A, Yuquan Rd., 100049 Beijing, China \\
\vs \no
   {\small Received [year] [month] [day]; accepted [year] [month] [day] }
}

\abstract{When  a satellite galaxy  falls into  a massive  dark matter
  halo, it suffers the dynamical friction force which drag it into the
  halo center  and finally it merger with the  central galaxy.  The time
  interval  between  entry  and  merger  is called  as  the  dynamical
  friction timescale ($T_{\rm df}$). Many studies have been dedicated to
  derive   $T_{\rm   df}$    using   analytical   models   or   N-body
  simulations. These  studies have obtained  qualitative agreements on
  how $T_{\rm  df}$ depends  on the orbit  parameters, and  mass ratio
  between  satellite   and  host   halo.  However,  there   are  still
  disagreements on the  accurate form of $T_{\rm df}$.  In this paper,
  we  present  a  semi-analytical  model to predict $T_{\rm df}$ and
  we focus  on interpreting the  discrepancies among different
  studies.  We find  that the treatment  of mass  loss from  satellite by
  tidal  stripping dominates  the behavior  of $T_{\rm  df}$.  We also
  identify other model parameters which affect the predicted $T_{\rm df}$.
 }

   \authorrunning{Gan et al. }
   \titlerunning{Dynamical Friction Timescale of Sinking Satellite}
   \maketitle

%%%%%%%%%%%%%%%%%%%%%%%%%%%%%%%%%%%%%%%%%%%%%%%%%%%%%%%%%%%%
\section{Introduction}
\label{sec:intro}

In the standard  cold dark matter (CDM) model,  structure (dark matter
halo) grows in  a hierarchical manner.  During the  merger of two dark
matter    haloes,     the    less    massive     one    becomes    the
satellite\footnote{When  we  refer  satellite,  we  mean  dark  matter
  subhalo, not  its luminous part  which is often called  as satellite
  galaxy.} (or subhalo) of the  more massive one ( host halo). The satellite
will  orbit  in  the  host  halo  and finally  merger  with  the  host
halo.  Halo  mergers play  an  important  role  in the  formation  and
evolution  of galaxies,  as  they can  significantly  affect the  star
formation    rate,    colors     and    morphology    of    galaxies
\citep[e.g.,][]{benson02,benson04,kang05,kazantzidis08}.     Therefore,
one inevitable  question about galaxy  formation and evolution  in the
CDM scenario is  to find out how long it takes  for the satellite to 
merge with the host halo.

Dynamical  friction  is  the  primary mechanism  which  decreases  the
orbital energy and  angular momentum of satellite, and  drag it to the
host halo center.  Description of dynamical friction was firstly given
by  \citet{chandrasekhar43},  who   derived  a  formula  of  dynamical
friction based on  the idealized case that a  rigid body moves through
an infinite,  homogeneous sea of  field particles.  For most  cases, the
satellite is moving  in a finite host halo,  and the dynamical friction
timescale  (\dft) of  satellite is  defined  as  the time  interval
between  entry   and  merger  with  the  host   center.   The  simple
application of  Chandrasekhar's formula  to drive \dft for  a rigid
satellite  is given by \citet[hereafter BT87]{binney87} and 
\citet[hereafter  LC93]{lacey93}, 
and these  formulas are  widely  used in  the  semi-analytical models for
galaxy               formation              and              evolution
\citep[e.g.,][]{kauffmann99,cole00,somerville99,neistein10}.      Early
study of \citet{navarro95} found  that the LC93 formula can accurately
match  their simulation  results. However,  the  simulation results of
\citet{springel01}  and \citet{kang05}  have indicated  that  the LC93
formula  underestimates the  merging timescale  and  overestimates the
merger rate as LC93 is only valid for a rigid object, not for a living
satellite in simulations.

For a  live satellite, one  needs to take  into account the  effect of
tidal   force which   leads  to the  mass  loss   from   satellite  and
redistribution of  mass inside  the satellite. Deriving  an analytical
formula of \dft for a live satellite is  nontrivial as one has
to  follow both  the orbit  and mass  evolution. 
\citet[hereafter C99]{colpi99}
firstly questioned the conclusion of  Navarro et al.  (1995), and they
found that  tidal stripping  can significantly increase  $T_{\rm df}$.
This   conclusion   was   recently   confirmed   by   
\citet[hereafter BK08]{boylankolchin08} and \citet[hereafter J08]{jiang08}
using high
resolution simulations.  BK08 and J08  both gave fitting  formulas for
$T_{\rm df}$, but with different dependence on orbit parameters. Their
results  differ by  a  factor up  to  2 for  eccentric orbits.   Using
semi-analytical   model   with   the   inclusion  of   tidal   effect,
\citet[hereafter T03]{taffoni03}  derived a fitting  formula for \dft.
However, their  results are not well tested  by simulations. Moreover,
the prediction of T03 is quantitatively inconsistent with the results
of  BK08 and  J08.

In  this  paper,  we  use a  semi-analytical model to study  \dft  of
satellite. Our main
motivation is neither to get a consistent result 
with simulation or other models, nor to derive a  reasonable \dft,
but to see how the model predictions
are affected  by various physical processes.  This  will tell us
which  process  dominates  the  predicted  \dft,  and  how  to
interpret  the discrepancies  among the previous  studies. 
Our  model is
based on \citet{taylor01} and \citet{zentner03}, but with a few 
modifications.
The paper is organized  as follows. In Section~\ref{sec:pre} we review
the    previous    results.      We    introduce    our    model    in
Section~\ref{sec:model},  and compare our  model predictions  with the
previous  work  in  Section~\ref{sec:result},  and  we  summarize  and
conclude briefly in Section~\ref{sec:conc}.

\section{The Previous Results}
\label{sec:pre}

\subsection{Set Up of  Initial Conditions}

The first step  of modelling the evolution of satellite is to  set its  
initial  conditions, including  the orbit  energy,  angular momentum and
 initial position. The  satellite is assumed to start  its orbit at the
virial  radius, $R_{\rm vir}$,  of the  host halo.  It has  an initial
orbital  energy equal  to that  of a  circular orbit  of  radius $\eta
R_{\rm vir}$,  and the initial specific angular  momentum of satellite
is parameterized as $j(0) =  \varepsilon j_{\rm c}$, where $j_{\rm c}$
is the specific angular momentum of the circular orbit mentioned above
and  $\varepsilon$  is the  orbital  circularity  (note  that $0  \leq
\varepsilon  \leq 1$).  In  the following,  we use  $\Rom$ to  denote the
initial  mass  ratio  between  the  host  and  satellite  halo,  i.e.,
$\Rom=M_h(0)/M_s(0)$.

\subsection{The Previous Results}
%%%%%%%%%%%%%%%%%%%%%%%%%%%%%%%%%%%%%%%%%%%%%%%%%%%
\begin{figure}[h!!!]
 \centering
 \includegraphics[width=\hsize]{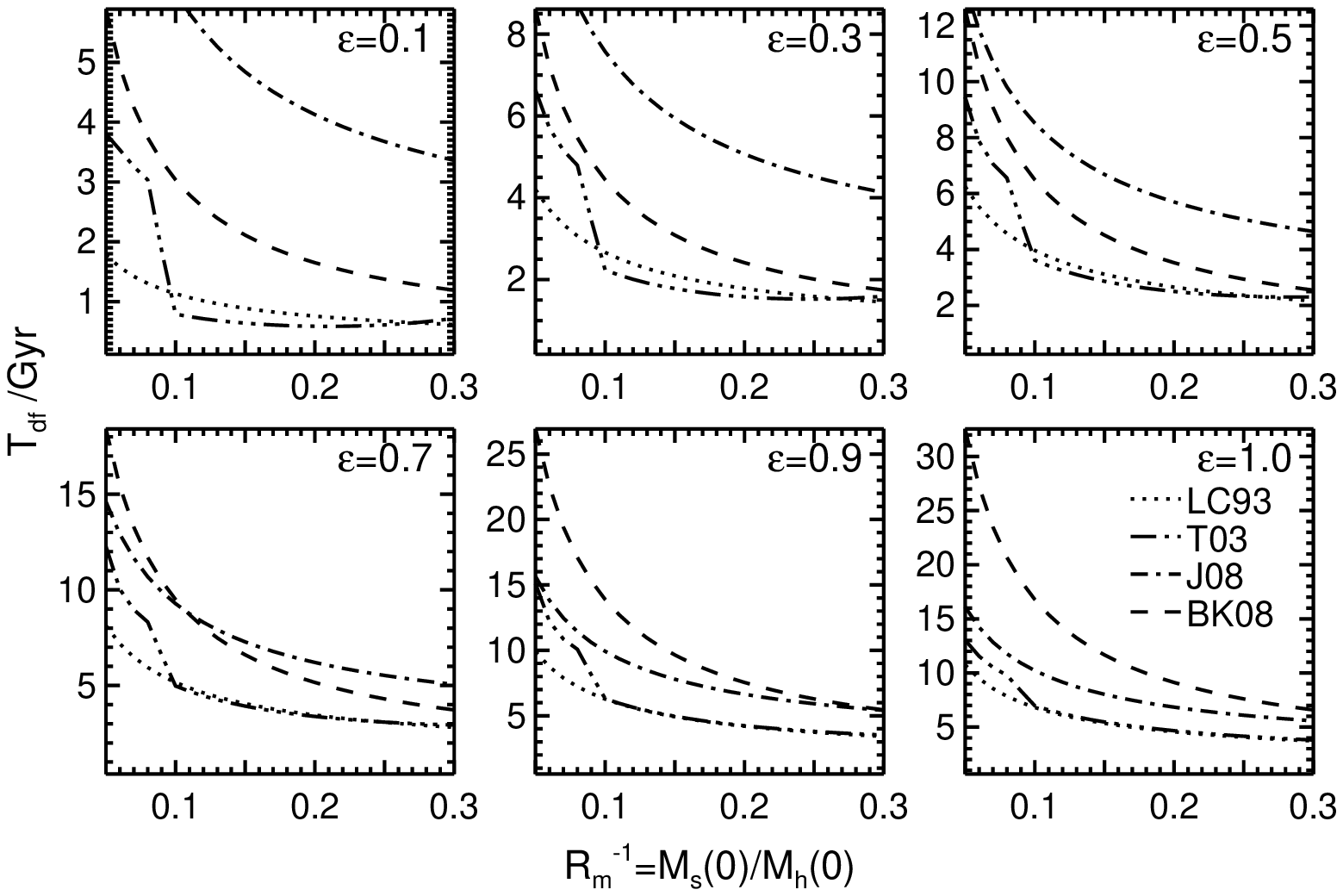}
 \caption{Dynamical friction timescale of sinking satellite predicted
   by the formulas of \citet[LC93]{lacey93}, \citet[T03]{taffoni03},
   \citet[BK08]{boylankolchin08} and \citet[J08]{jiang08}.
   The six panels show the dependence on the orbital
   circularity $\varepsilon$, and each panel shows the
   dependence on the initial mass ratio between the satellite
   and host halo. In the results of LC93, we adopt
   $\ln\Lambda=\ln(1+\Rom)$, as used by T03, BK08 and J08.
   $\eta=1.0$ is used in all cases. }
 \label{fig:tprior}
\end{figure}
%%%%%%%%%%%%%%%%%%%%%%%%%%%%%%%%%%%%%%%%%%%%%%%%%%%%%%

Here  we  briefly  review  the previous  studies  on  $T_{\rm  df}$ from 
analytical  models or  N-body simulations.   Using  the Chandrasekhar's
formula, BT87  derived an expression of  \dft for satellite
starting with circular orbit in an isothermal distributed host halo as
\begin{equation}
\label{eq:BT87}
T_{\rm df, BT87}=\frac{1.17}{\ln \Lambda}\Rom\tau_{\rm dyn}
\end{equation}
where $\tau_{\rm  dyn}$ is  the dynamical time  $R_{\rm vir}/V_{\rm vir}$,
and \lld is the \CL.

Taking  into account  the dependence  on the  orbital  circularity, 
LC93 obtained that
\begin{equation}
\label{eq:LC93}
  T_{\rm df, LC93}=\frac{\varepsilon^{0.78}}{0.855}
  \frac{\Rom}{\ln \Lambda} \eta^2 \tau_{\rm dyn} \;,
\end{equation}
Note that in the above two equations, the satellite is treated as a rigid 
object without mass loss. 

With help of N-body simulation, C99 derived \dft for a live satellite as
\begin{equation}
\label{eq:C99}
  T_{\rm df, C99}=1.2\varepsilon^{0.4}\frac{\Rom}{f_m\ln \Lambda}
  \eta^2 \tau_{\rm dyn} \;,
\end{equation}
where $f_m$ refers to the remaining fraction of satellite mass due to 
tidal stripping.  Note that C99 only considers minor mergers. It's
difficult to use this formula  as the \dft depends on the 
presumed value for $f_{m}$. 

%C99 use 0.2 as a fiducial value for $f_{m}$.

Using a  semi-analytical model, T03 derived their  fitting formulas for
\dft,  and   they were  updated   by  \citet{monaco07}.  Their model have
incorporated the effect of tides,  but they ignore this effect for the
large satellite (with  mass $\Rin >0.1$).  Here we omit the complex
formula of T03.

Using smoothed-particles hydrodynamical  simulation with
gas cooling and star formation in a cosmological context, 
J08 fitted their results with \dft as:
\begin{equation}
\label{eq:J08}
  T_{\rm df, J08}=\frac{0.9\varepsilon^{0.47}+0.6}{0.855}
  \frac{\Rom}{\ln(1+\Rom)} \sqrt{\eta} \tau_{\rm dyn} \;.
\end{equation}

BK08  considered controlled  N-body simulations for two halo mergers. 
They gave the fitting formula of \dft as
\begin{equation}
\label{eq:BK08}
  T_{\rm df, BK08}=0.216 e^{1.9\varepsilon} \frac{\Rom^{1.3}}
  {\ln(1+\Rom)} \eta \tau_{\rm dyn} \;.
\end{equation}

In Figure~\ref{fig:tprior}  we show the \dft as  function of satellite
mass and  orbital circularity\footnote{Throughout this  paper, we keep
  the  orbital energy  fixed as  $\eta=1.0$ to  reduce the  free model
  parameters.}   predicted by  LC93, T03,  J08 and  BK08.  For  a full
comparison with  other results, we  choose $\ln\Lambda=\ln(1+\Rom)$ in
the formula  of LC93.  It  can be seen  that all results show  a clear
trend  that \dft  decreases with  the increasing  satellite  mass, and
increases with the orbital  angular momentum and energy.  However, the
discrepancies  among  different  studies  are still  remarkable.   For
example, the results  of BK08 and J08 are  longer than that
of T03 and LC93. T03 agrees  well with LC93 for large satellite ($\Rin
> 0.1$),  but disagrees  for  small satellite.   The  results of  BK08
exhibits a steeper dependence  on $\varepsilon$ than other results.

\section{Modelling the Sinking Satellite}
\label{sec:model}

This section  describes the dynamical evolution of  satellite based on
the model of \citet{taylor01,zentner03}.  In section~\ref{subsec:halo}
we introduce the model for  the mass distribution of dark matter halo.
Then we describe the physical processes governing the orbital and mass
evolution   of  satellite.    These  process   can   be  independently
implemented into the model, which  allows us to investigate the effect
of any specific process by tune its free parameter.

\subsection{Halo Properties}
\label{subsec:halo}

The dark matter halo is  a gravitational self-bound system. We express
the size of halo in terms  of its virial mass $M_{\rm vir}$ and virial
radius $R_{\rm vir}$, which is  defined as the radius within which the
mean  mass density of  the halo  is $200$  times the  critical density
($\rho_c$) of the universe at $z=0$ \citep[e.g.,][]{mo98}.  The Hubble
constant  is  adopted to  be  $H_0=100h{\rm km~s^{-1}~Mpc^{-1}}$  with
$h=0.7$ (BK08). The dynamical timescale can be described as
\begin{equation}
 \tau_{\rm dyn}=\frac{R_{\rm vir}}{V_{\rm vir}}=\left(\frac{R_{\rm vir}^3}
 {GM_{\rm vir}}\right)^{1/2}=0.1H_0^{-1}\simeq 1.40 {\rm Gyr} \;,
\end{equation}
where ${V_{\rm vir}}$ is the virial velocity of a halo.

For simplicity, the dark matter halo is usually treated  as a spherically  
symmetric system, and a simple formalism for the halo density profile
is the profile of singular isothermal sphere (hereafter, ISO profile), 
which can be described by \citep[e.g.,][]{mo98}
\begin{equation}
 \rho(r)=\frac{V_{\rm vir}^2}{4\pi G r^2} \;,
\end{equation}
and
\begin{equation}
 M(<r)=\frac{V_{\rm vir}^2}{G} r \;.
\end{equation}

As measured by N-body  simulations, the halo density profile can be
well described by the NFW profile (Navarro et al. 1997):
\begin{equation}
 \rho(r)=\frac{\delta_0 \rho_c}{(r/r_s)(1+r/r_s)^2} \;,
\end{equation}
with  $r_s$  the  scale  radius,  and  $\delta_0$  the  characteristic
overdensity. From  the definition  of virial radius,  we can  find the
characteristic   overdensity  that   $\delta_0=200c^3/[3g(c)]$,  where
$c=R_{\rm   vir}/r_s$  is  the   halo  concentration   parameter,  and
$g(x)=\ln(1+x)-x/(1+x)$. For the NFW profile, the halo mass enclosed a
radius $r$ is
\begin{equation}
 M(<r)=M_{\rm vir}\frac{g(r/r_s)}{g(c)} \;.
\end{equation}
The halo concentration is tightly correlated to its mass, and we use the
median relation of $c \sim M$ as measured by \citet{neto07}:
\begin{equation}
\label{eq:cm}
c(M)=4.67\left[\frac{M}{10^{14}h^{-1}M_{\odot}}\right]^{-0.11} \;.
\end{equation}

Note that there are still debates existing in the inner shape of the
NFW profile \citep[e.g.,][]{fukushige01,navarro04,stoehr06,springel08}.
Varying the shape of NFW profile or using other halo profiles
[e.g., ISO profile; Hernquist porfile \citep{hernquist90}] may derive a
different \dft.  However, the simulation of BK08 indicated that using a 
different halo profile
had a change in \dft  of only $5\%$ (see BK08 for more
details).   

Except for Section~\ref{subsec:examine} where the ISO profile is
adopted to compare the model predictions with the analytical results of 
LC93, we use the NFW profile in other studies of this paper.
When the tidal effects are considered, the satellite halo
has a NFW  profile  at  the time  of entering  ($t=0$), and this  
profile is  subsequently modified  due to 
tidal heating, as described in Section~\ref{subsec:heating}.

In our studies, we select the host halo mass as $10^{12}M_{\odot}$, which
is  the typical  mass used  to derive  the \dft  (BK08, J08,
C99).  We have  also  tested that  the  predicted \dft has  a
negligible effect on the host halo mass once the mass ratio $\Rom$ is
fixed.

\subsection{Dynamical Friction}
\label{subsec:df}

The satellite will sink into the halo center by the dynamical friction
force  which is caused  by the  gravitational interaction  between the
satellite and the  background `field' particles that make  up the host
halo  (for a  complete description,  see BT87).   This  effect was
first discussed by \citet{chandrasekhar43}, and the force generated by
the filed particles is  known as the Chandrasekhar dynamical friction.
By assuming  that the  field particles  follow a  locally Maxwellian
velocity  distribution, BT87 gave the formula of dynamical friction as
\begin{equation}
\label{eq:f_df}
{\bf F}_{\rm df} = -4\pi G^2 M_s^2
 \ln\Lambda \, \rho(r) \left[{\rm erf}(X)-{2X \over \sqrt{\pi}}
 e^{-X^2}\right] {{\bf v}_{\rm orb}\over v_{\rm orb}^3} \;,
\end{equation}
where $v_{\rm orb}$ is the orbital velocity of the satellite, and
$X=v_{\rm orb}/[\sqrt{2}\sigma(r)]$ with $\sigma(r)$ the local,
one-dimensional velocity dispersion of the host halo at radius $r$,
which can be solved from the Jeans equation \citep[BT87,][]{cole96}. 
For ISO profile, $\sigma(r)\equiv V_{\rm vir}/\sqrt{2}$; for NFW
profile, we use the fitting formula of $\sigma(r)$ from
\citet{zentner03}. We choose the \CL $\ln\Lambda=\ln(1+\Rom)$,
as used by T03, J08 and BK08.

The Equation~(\ref{eq:f_df}) was derived with the
idealized assumption that the velocity distribution of the dark matter 
particles is Maxwellian and isotropic. Although there are
debates on whether this assumption is reasonable
\citep[e.g.,][]{manrique03,williams04,salvadorsole05,bellovary08}, in this
paper, we follow most authors (e.g., LC93; C99; T03; Zentener \& Bullock 
2003; Fellhauer \& Lin 2007; BK08) to adopt the Maxwellian and isotropic 
velocity distribution. There are also simulations showing that this 
assumption is a good approximation 
\citep[e.g.,][]{cole96,sheth96,seto98,kang02,hayashi03}.

\subsection{Tidal Mass Stripping}
\label{subsec:stripping}

For a live satellite, the tidal force from the host halo will strip 
its mass. The tidal radius, $r_{\rm t}$, is the distance from the
center of satellite  to the radius where the external
differential force from the host halo exceeds the binding force of
the satellite. The tidal radius can be simply solved from the
following equation \citep{vonhoerner57,king62,taylor01}:
\begin{equation}
 \label{eq:rt}
 r_{\rm t}^3 = \frac{G M_s(<r_{\rm t})}
{\omega^2 + G\left[ 2M_h(<r)/r^3 - 4\pi\rho_h(r) \right]} \;,
\end{equation}
with $\omega$ the angular speed of the satellite and $\rho_h(r)$ the
density profile of the host halo. 
The mass outside $r_{\rm t}$ becomes unbound and is stripped gradually. 
\citet{taylor01} suggested the unbound mass to be stripped at the rate 
that
\begin{equation}
 \label{eq:strip_tb01}
 {{\rm d}M_s \over {\rm d}t}=-\frac{M_s(>r_{\rm t})}{T_{orb}} \;,
\end{equation}
with $T_{orb}$ the instantaneous orbital period (i.e.,
$T_{orb}=2\pi /\omega$), which is assumed as the mass stripping
timescale.

There are some uncertainties in the above mechanisms of mass stripping.
(i) The tidal radius cannot be characterized by a single radius, as
the zero-velocity surface (the surface defined by the tidal radius, see 
BT87) is not spherical. 
(ii) The perturbation of particles within the satellite may 
lead to the scatter in $\omega$, and the zero-velocity surface is actually
a shell of `non-zero' thickness, while this effect is ignored in
Equation~(\ref{eq:rt}). So the solution of Equation~(\ref{eq:rt}) is only
an approximation for the tidal radius. 
(iii) The stripped mass from a satellite still remain in the vicinity of 
the satellite, and the interaction between
the stripped and unstripped mass will perturb the satellite orbits and
affect the mass loss \citep[e.g.,][]{fellhauer07}.

Owing to these uncertainties,  numerical simulations have debated on how 
fast the unbound mass is stripped from the 
satellite. \citet{zentner05} and \citet{diemand07} found a stripping
timescale $3.5$ and $6$ times shorter than $T_{orb}$, respectively. It
was also pointed out that the stripping timescale is dependent on the
satellite internal structures \citep{kazantzidis04,kampakoglou07}. In
general, the mass loss rate can be described using a free parameter
$\alpha$ as:
\begin{equation}
 \label{eq:strip}
 {{\rm d}M_s \over {\rm d}t}=-\alpha \frac{M_s(>r_{\rm t})}{T_{orb}} \;,
\end{equation}
where $\alpha$  describes the efficiency of tidal stripping.  In
Section~\ref{subsec:alpha} we will show how the \dft depends on
$\alpha$.

\subsection{Tidal Heating}
\label{subsec:heating}
During the pericentric passage of satellite orbits, the gravitational
field changes rapidly, and this induces a gravitational shock that can
add energy to the satellite \citep[e.g.,][]{gnedin97,gnedin99}. This
effect is called the tidal heating.
It has been found from $N$-body simulations
\citep[e.g.,][]{hayashi03,kravtsov04} that tidal heating will expand
the satellite and reduce its inner mass profile. \citet{hayashi03}
introduced a modified NFW profile to describe the density
distribution of a tidally heated satellite according to
\begin{equation}
\label{eq:heating}
\rho(r) = \frac{f_t}{1+(r/r_{te})^3} \rho_{NFW}(r) \;,
\end{equation}
where
\begin{equation}
\label{eq:ft}
\lg f_t=-0.007+0.35 x_m + 0.39 x_m^2 + 0.23 x_m^3 \;,
\end{equation}
and
\begin{equation}
\label{eq:rte}
\lg \frac{r_{te}}{r_s} = 1.02 + 1.38 x_m + 0.37 x_m^2 \;.
\end{equation}
In Equation~(\ref{eq:heating}), $\rho_{NFW}(r)$ is the original NFW
density profile of the satellite at the time of entering ($t=0$), $f_t$
describes the reduction in the central density of the satellite, and
$r_{te}$ is the `effective' tidal radius that describes the outer cutoff
imposed by the tides. In Equation~(\ref{eq:ft}) and (\ref{eq:rte}),
$x_m = \lg[M_s(t)/M_s(0)]$ is the logarithm of the remaining fraction of
satellite mass, and $r_s$ is the scale radius of the satellite with
NFW profile at $t=0$. As shown by \citet{hayashi03}, $f_t$ and $r_{te}$
are well fitted by the function of $x_m$. Both $f_t$ and $r_{te}$ decrease
with time while a satellite is losing mass.

\subsection{Orbital Evolution}
\label{subsec:orbit}

Here we present explicitly the equations to solve the orbit 
[${\bf x}(r, \theta)$] of the satellite  under gravity and the dynamical 
friction. The equation of motion for
the satellite is given by
\begin{equation}
 \label{eom}
 {{\rm d}^2{\bf x} \over {\rm d}t^2} = -{G M_h(<r) \over r^3}\,{\bf r}
  + {{\bf F}_{\rm df} \over M_s}
\end{equation}
with $M_h(<r)$  the mass of  the host halo  inside of radius  $r$, and
${\bf   F}_{\rm   df}$  the   dynamical   friction   force  given   by
Equation~(\ref{eq:f_df}). The  orbital energy and  angular momentum of
the satellite will decay due to the dynamical friction as it is always
opposite to the direction of motion.  We define satellite to be merged
with host center when it loses all its angular momentum, and \dft is the
time interval between  accretion   and    merger    \footnote{Some
  \citep[e.g.,][]{kravtsov04,zentner05}   define  satellite   to  be
  merged with  the host halo when  its distance to the  host center is
  less than a fiducial radius.  We find that different definitions have
  no significant  effects.} (as used  also by BK08).  The  equation of
motion and Equation~(\ref{eq:strip})  are solved using the fifth-order
Cash-Karp Runga-Kutta method, in which an adaptive step-size control is
embedded.

\section{Results}
\label{sec:result}
%In this Section, we show the predicted \dft from our model.

\subsection{Examination on a Rigid Satellite}
\label{subsec:examine}
%%%%%%%%%%%%%%%%%%%%%%%%%%%%%%%%%%%%%%%%%%%%%%%%%%%
\begin{figure}[h!!!]
 \centering
 \includegraphics[width=\hsize]{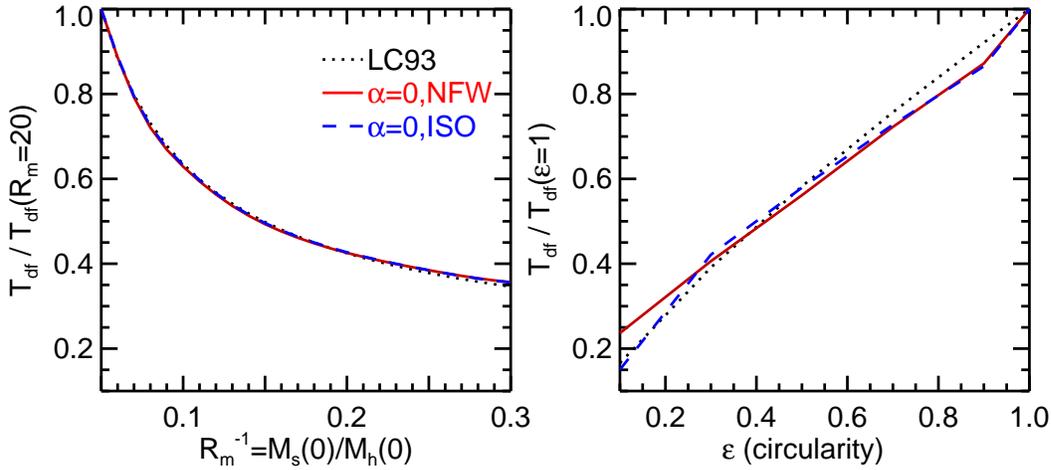}
 \caption{The dependence of \dft on satellite mass (left panel) and orbital
 circularity (right panel) for a rigid satellite, where \dft is normalized
 to its value  when $\Rom=20$ and $\varepsilon=1$, respectively.
 The model with  $\alpha=0$  means that a rigid satellite is considered. 
 The results in red solid lines are computed with NFW profile, while the
 blue dashed lines show the results with ISO profile. 
 Both model predictions  match well with LC93's (balck dotted). }
 \label{fig:tdep0}
\end{figure}
%%%%%%%%%%%%%%%%%%%%%%%%%%%%%%%%%%%%%%%%%%%%%%%%%%%%
%%%%%%%%%%%%%%%%%%%%%%%%%%%%%%%%%%%%%%%%%%%%%%%%%%%%
\begin{figure}[h!!!]
 \centering
 \includegraphics[width=\hsize]{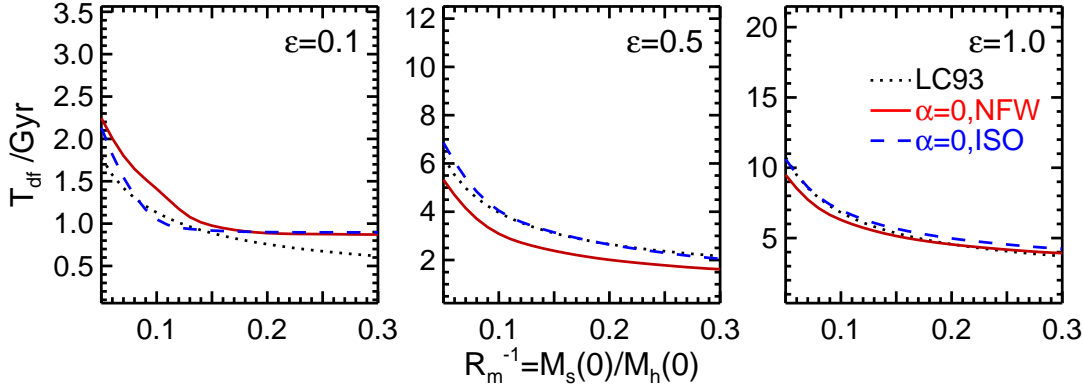}
 \caption{Comparison of \dft between our model ($\alpha=0$)
 and LC93 (dotted) for a rigid satellite.  The \dft with NFW profile
 (red solid) and  ISO profile (blue dashed) both agree well with LC93's 
 prediction, and they agree for 
 all orbital circularity, although only $\varepsilon$ with $0.1, 0.5, 1.0$
 are given here.}
 \label{fig:tmlr0}
\end{figure}
%%%%%%%%%%%%%%%%%%%%%%%%%%%%%%%%%%%%%%%%%%%%%%%%%%%%% 

Firstly we validate  our model by comparing the  predicted \dft with the
LC93  result  for a  rigid  satellite. LC93 derived \dft using
Equation~(\ref{eq:f_df}) and ISO profile for the host halo. In  our  model, 
we simply set $\alpha=0$ to `close' the tidal stripping and tidal heating 
effect, and we model the host halo with both NFW profile and ISO profile.

In Figure~\ref{fig:tdep0} we show the \dft as a function of $\Rin$ and
$\varepsilon$ for a rigid satellite, with \dft normalized to its value
when  $\Rom=20$ and $\varepsilon=1$,  respectively. As  indicated, our
results in NFW (red solid) and ISO (blue dashed) model both have the  same dependences as predicted 
by  LC93. On the other 
hand, the amplitudes  of \dft from the models also  agree well with the
results of LC93, which  is demonstrated in Figure~\ref{fig:tmlr0}. The
difference  resulted by varying halo profile are small and negilible,
which is also concluded by BK08.

%between  our model and LC93 is from  that we
%employ  a NFW  halo profile,  while  LC93 used  a singular  isothermal
%profile.

\subsection{Dependence on Tidal Stripping Efficiency $\alpha$}
\label{subsec:alpha}
%%%%%%%%%%%%%%%%%%%%%%%%%%%%%%%%%%%%%%%%%%%%%%%%%%%
\begin{figure}[h!!!]
 \centering
 \includegraphics[width=\hsize]{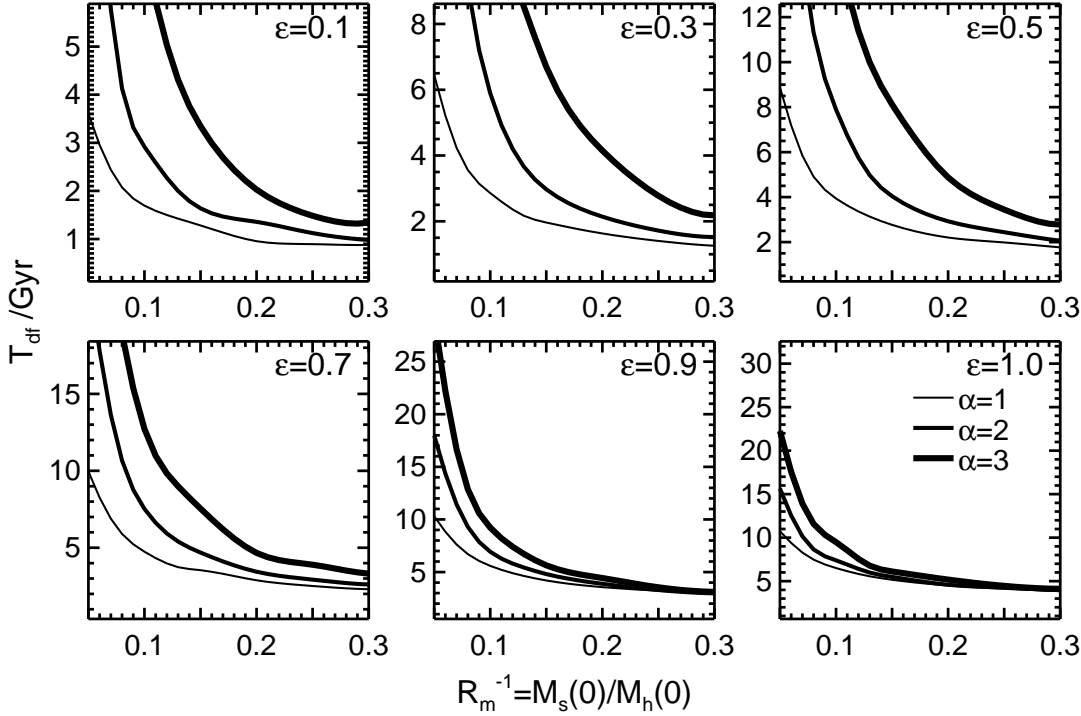}
 \caption{The dynamical friction timescale (\dft) predicted by our model.
 The lines with increasing thickness  show the
 effects of tidal  stripping efficiency ($\alpha=1, 2, 3$). The \dft is
 a strong function of $\alpha$.  }
 \label{fig:tmlr}
\end{figure}
%%%%%%%%%%%%%%%%%%%%%%%%%%%%%%%%%%%%%%%%%%%%%%%%%%%
%%%%%%%%%%%%%%%%%%%%%%%%%%%%%%%%%%%%%%%%%%%%%%%%%%%
\begin{figure}[h!!!]
 \centering
 \includegraphics[width=\hsize]{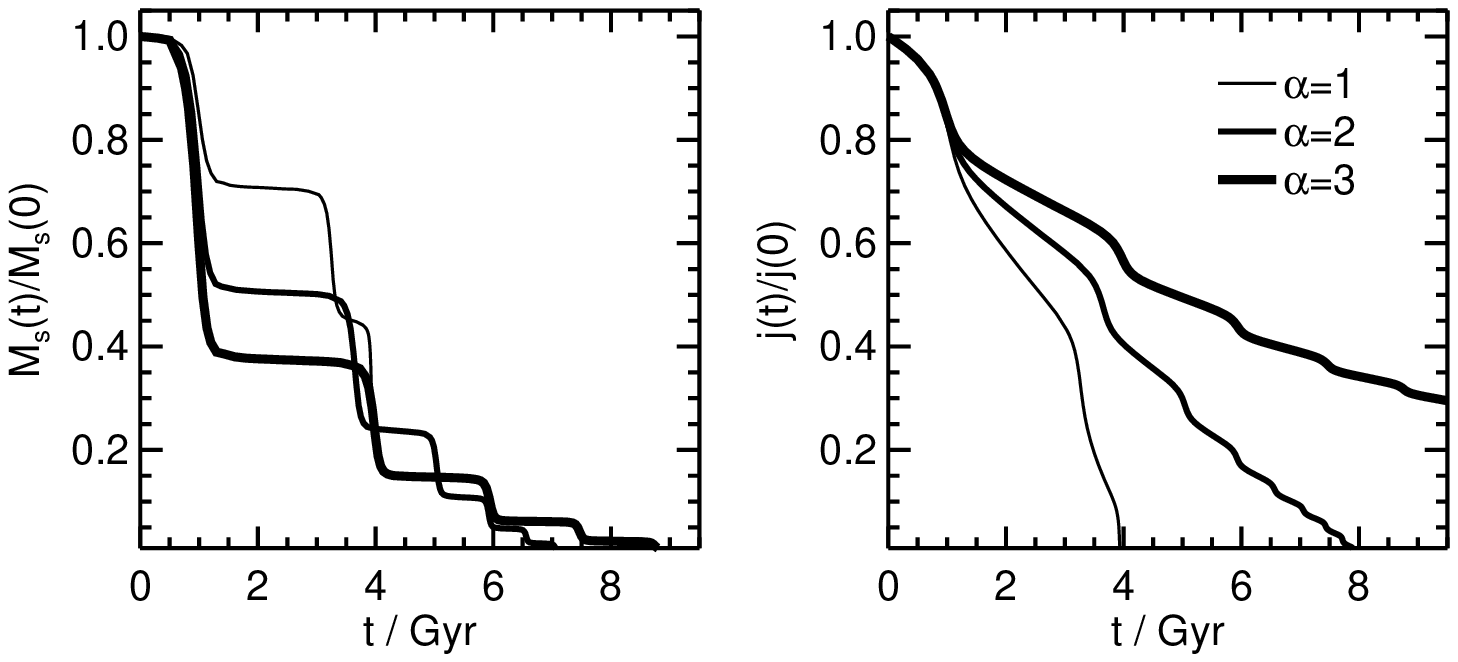}
 \caption{The evolution of satellite mass and specific angular momentum
 (both are normalized to their initial value)
 as function of tidal stripping efficiency $\alpha$. The initial
 conditions are that $\Rom=10$, $\varepsilon=0.5$, $\eta=1.0$. Strong
 tidal effects reduce the amplitude of dynamical friction and decelerate
 the loss of angular momentum. }
 \label{fig:mj}
\end{figure}
%%%%%%%%%%%%%%%%%%%%%%%%%%%%%%%%%%%%%%%%%%%%%%%%%%%

In this  section we  study the effects  of tidal  stripping efficiency
($\alpha$). In  Figure~\ref{fig:tmlr} we show the  predicted \dft with
different values of $\alpha$.  A larger value
of $\alpha$ corresponds to a stronger tidal field or a rapid mass loss
from the satellite.  The results show a remarkable trend that the \dft
is increased when the tidal field becomes stronger.  The reason can be
seen from  Figure~\ref{fig:mj} which shows the  evolution of satellite
mass and  specific angular momentum with dependence  on $\alpha$.  The
initial  conditions  are  set  as  $\Rom=10$,  $\varepsilon=0.5$,  and
$\eta=1.0$.  The  left panel  shows that a  stronger tidal  field will
induce  more mass loss  from the  satellite, and  this effect  is more
distinct at the beginning.  As seen from Equation~(\ref{eq:f_df}), the
amplitude of dynamical friction has a strong dependence on the mass of
satellite ($F_{\rm df} \propto  M_s^2$). So a stronger tidal stripping
will lead to  a slower decay of satellite  angular momentum and result
in a longer dynamical friction  timescale, as shown in the right panel
of Figure~\ref{fig:mj}.

%%%%%%%%%%%%%%%%%%%%%%%%%%%%%%%%%%%%%%%%%%%%%%%%%%%
\begin{figure}[h!!!]
 \centering
 \includegraphics[width=\hsize]{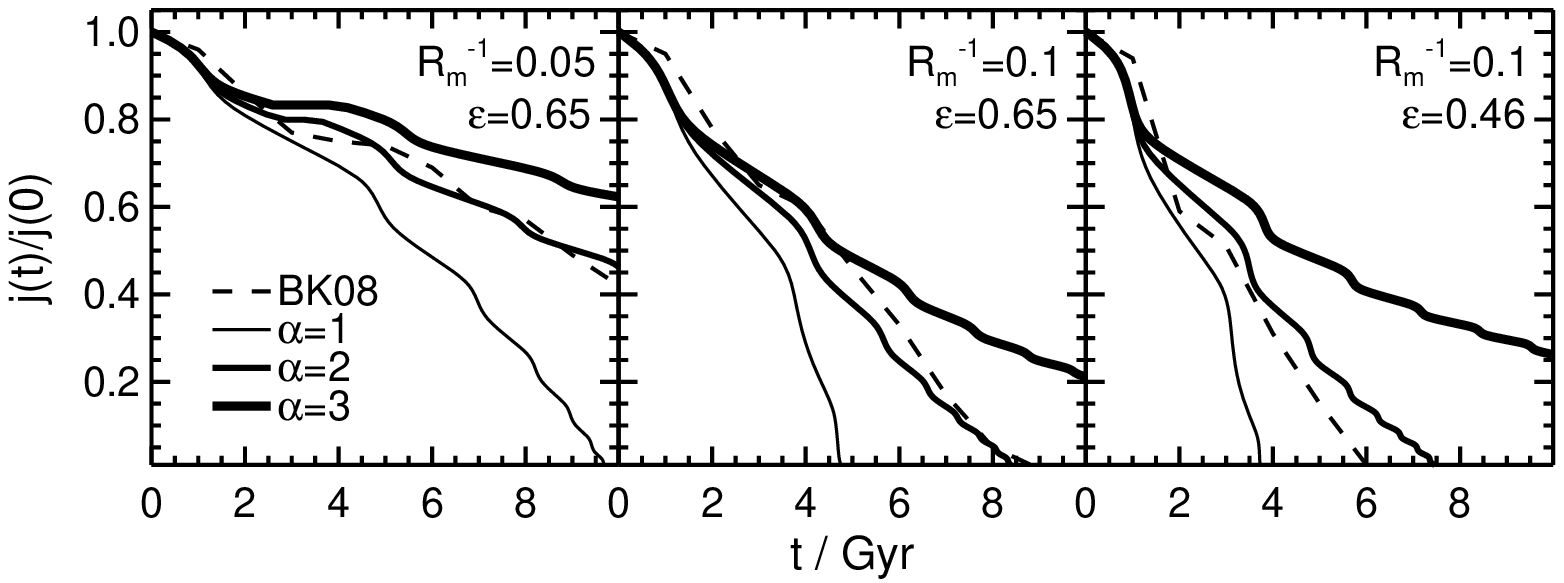}
 \caption{Comparison of the evolution of satellite specific angular
 momentum between our model and BK08, with three cases of initial masses
 and orbits as indicated. The solid lines with increasing
 thickness are the model results with $\alpha=1, 2, 3$, while the dashed
 line is the result from Figure~1 of BK08. The tidal stripping
 efficiency in the simulation of BK08 should be stronger than that in a model
 with $\alpha=1$.  }
 \label{fig:bk}
\end{figure}
%%%%%%%%%%%%%%%%%%%%%%%%%%%%%%%%%%%%%%%%%%%%%%%%%%%

As  shown  in Figure~\ref{fig:tprior},  the  predicted  \dft from  the
previous results  disagree  with  each other  quantitatively.
We believe  that the main discrepancy  is resulted in  the treatment of
tidal stripping, and we discuss it in more details in the following.
\begin{itemize}
\item  T03 ignored the  tidal effects for  massive satellite  (with mass
  $\Rin >0.1$), and so their  \dft are consistent with LC93's. But T03
  predicted a longer \dft for low-mass satellite which suffers from tidal
  stripping.

\item  The \dft  inferred by J08 and BK08 are longer  than that of T03.
  This is  because T03  adopted a tidal  stripping efficiency  that is
  different   from   those    in   N-body   simulations.    T03 also used
  Equation~(\ref{eq:strip})  to   describe  the  mass   loss,  but  with
  $\alpha=1.0$ which is too low.  As shown by \citet{zentner05}, a
  higher value that  $\alpha=3.5$  is  required  to  better fit  the  
  satellite  mass function  from  simulations \citep[also see][]{gan10}. 
  A  higher  value  of  $\alpha$ is  also
  favored from  Figure~\ref{fig:bk} where we compare  the evolution of
  satellite specific angular momentum  from our model (solid lines) with
  the  simulation  results  of  BK08  (dashed  lines).  We  find  that
  $\alpha=2$ can  better match the simulation results.  Thus the lower
  value of  $\alpha$ used  by T03 explains  why they obtained  a lower
  $T_{\rm df}$.

\item  The \dft of J08 is longer than that of BK08 for  eccentric orbit
 (i.e., low $\varepsilon$)\footnote{The results of BK08 and J08 also 
 differ for small satellite with large $\varepsilon$, of which the \dft, 
 however, are extrapolated by their formulas and exceed the Hubble time.}.
 The simulation of J08 includes the process
 of gas cooling and star formation. The halo of a satellite is expected
 to contract in response to the cooling of gas
 \citep[e.g.,][]{gnedin04,abadi10}. During the
 pericentric passage, the satellite with halo contraction is resistant to
 the strong tidal field, and will survive for a longer time
 \citep[e.g.,][]{weinberg08,dolag09}. Instead BK08
 performs a higher resolution simulation, in which the satellite
 can avoid the artificial mass loss due to the numerical effects. So the
 satellite will deposit more mass in the eccentric orbit and
 suffer stronger dynamical friction.
\end{itemize}

\subsection{Dependence on Orbital Circularity $\varepsilon$}
%%%%%%%%%%%%%%%%%%%%%%%%%%%%%%%%%%%%%%%%%%%%%%%%%%%
\begin{figure}[h!!!]
 \centering
 \includegraphics[width=0.6\hsize]{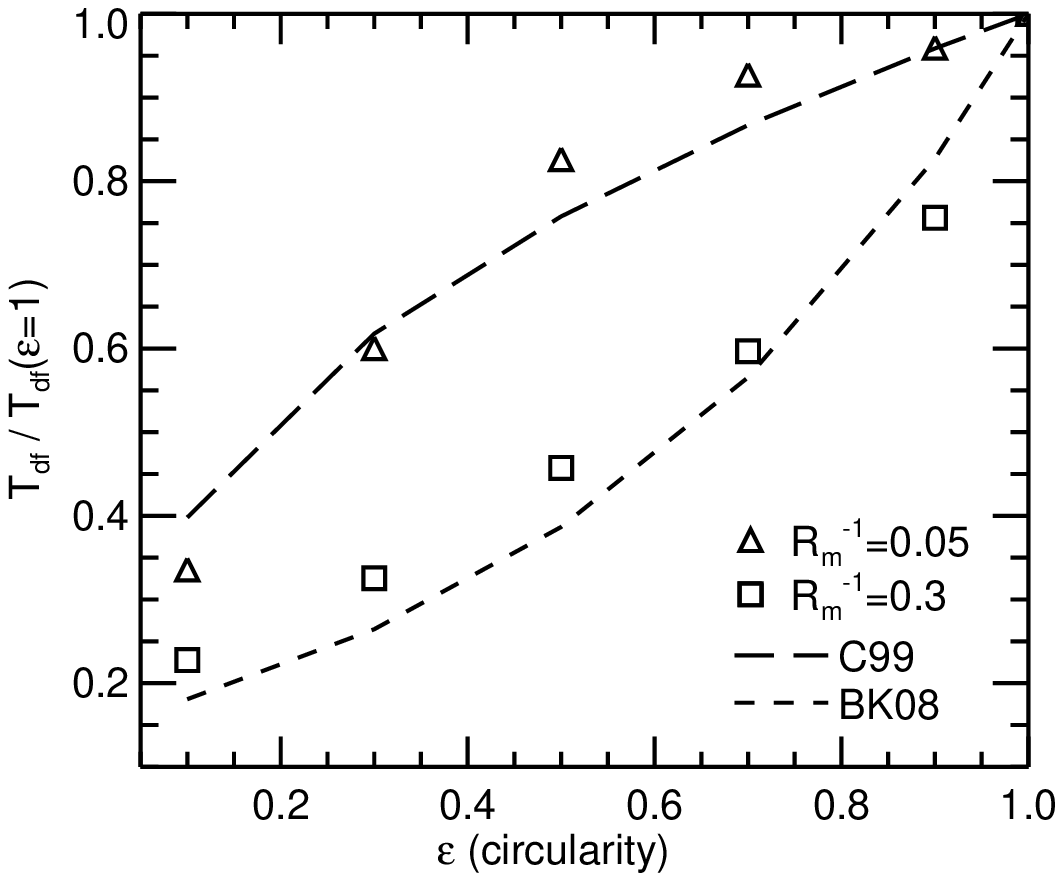}
 \caption{The dependence of \dft on  orbital circularity for a live
 satellite ($\alpha=1$). For a minor merger (triangle), with
 $\Rin =0.05$, the
 result shows a power-law dependence, which is similar to that of C99
 (long-dashed),  while for a major merger (square), with $\Rin =0.3$,
 it  indicates an  exponential dependence, which is close to that of
 BK08  (short-dashed).}
 \label{fig:tdep1}
\end{figure}
%%%%%%%%%%%%%%%%%%%%%%%%%%%%%%%%%%%%%%%%%%%%%%%%%%%%

The previous results showed similar  dependence of \dft on the initial
satellite  mass,  but  very   different  dependences  on  the  orbital
circularity [Equations~(\ref{eq:LC93})-(\ref{eq:BK08})].  For example,
BK08  found an  exponential  dependence  of \dft  on  the the  orbital
circularity,  while  others  found  a power-law  dependence.  Here  we
investigate this  problem using our model with  $\alpha=1$. We compute
the  \dft as a  function of  $\varepsilon$ for  a minor  merger ($\Rin
=0.05$)   and   a   major   merger   ($\Rin  =0.3$),   as   shown   in
Figure~\ref{fig:tdep1}. We  find the  dependence for the  minor merger
can be fitted to a  power law, $T_{\rm df} \propto \varepsilon^{0.4}$,
as  predicted  by  C99   (long-dashed).  For  the  major  merger,  the
dependence is close  to the result of BK08,  who found the exponential
law  that  $T_{\rm  df} \propto \exp (1.9\varepsilon)$. It is  not  a
surprise as C99 only considers minor mergers while BK08 has more
samples for  the major mergers. Thus  we argue that  the dependence on
orbital circularity  is mainly determined by the  distribution of mass
ratio between the satellite and host halo.

\section{Conclusion and Discussion}
\label{sec:conc}

In this paper,  we study the dynamical friction  timescale (\dft) of a
sinking satellite into a host halo.  Previous results using analytical
models or simulations generally agree that the \dft is correlated with
the  mass,  orbital  circularity  and  energy of  the  satellite,  but
disagree  on the  amplitude  of \dft  and  the dependence  of \dft  on
orbital  circularity.  It  was   unclear  what  contributes  to  these
discrepancies among different studies.

Aiming  at  interpreting  these   different  dependences,  we  use  a
semi-analytical model similar to that of \citet{taylor01} and 
\citet{zentner03} to derive the \dft.   
Our model considers the main  physical processes governing
the evolution of satellite: dynamical friction, tidal stripping, tidal
heating and  merger. All these process are  independently described by
free parameters,  and it  allows us to  investigate the  dependence of
$T_{\rm df}$ on any process.

Firstly, we apply our model to  a rigid satellite by `turning off' the
tidal  stripping  and tidal  heating  (i.e.,  $\alpha=0$).  The  model
predictions agree well with the LC93's result on the amplitude of \dft
and its  dependences on satellite mass and  orbital circularity.  Then
we study the dependence of  \dft on the tidal stripping efficiency. We
find that the  \dft depends strongly on $\alpha$,  with the trend that
the \dft  increases with increasing $\alpha$.  A  higher $\alpha$ leads
to rapid  loss of  mass from satellite,  than decreases  the dynamical
friction force. Thus  this results in a slower  decay of angular momentum
and a longer  \dft. We believe that the main  reason for the diversity
of previous result is the treatment of tidal stripping.

We  also study  the  dependence  of \dft  on  the orbital  circularity
($\varepsilon$). We find that for low mass-ratio mergers ($M_s/M_h<0.1$),
\dft is  a power  law of  orbital  circularity. While  for massive  
mergers ($M_s/M_h>0.1$),  the  dependence of \dft on orbital circularity 
is expoential. Thus we argue that the dependence on $\varepsilon$ obtained
by different  studies is  determined by their samples, in  which the mass
ratio between satellite and host halo is crucial.

In this paper, we do not model the effects of baryon,
as it is difficult to include the physical processes governing  
galaxy formation, and it is still not clear how dark matter halo will  
respond to the baryon at the  host halo center.

The major effect of baryon is to modify the density profile of dark matter 
halo. There are still debates about how the baryon will change the central 
concentration  of halo. Some found that central density  increases
\citep[e.g.,][]{blumenthal86,gnedin04}, but some disagreed with it. 
\citet{gnedin04} found that the halo will become more concentrated as 
baryons condense in the radiative cooling, and the contraction of halo is 
dependent on the amount of baryon. While \citet{abadi10} found that the 
response of halo contraction depends not only on how much baryon mass has 
been deposited  by the halo, but also on the mode of its deposition 
\citep[also see][]{tissera10}. They showed that strong feedback by
supernovae can significantly decrease the central density of halo 
\citep[also see][]{pedrosa09,governato10}. 
The variation of \dft is about $20\%$ when $c_{sat}/c_{host}$ changes 
between $1$ and $2$ (T03; BK08).

There are also some studies
showing that the dark matter haloes have constant density cores 
\citep[e.g.,][]{gilmore07,deblok08,kuziodenaray09,gebhardt09,hernandez10},
which can signficantly suppress the effect of dynamical friction
\citep[e.g.,][]{sanchezsalcedo06,inoue09}. Howerer, the typical size of the 
constant density core in the dark matter halo is usually less than $1$~kpc
\citep[e.g.,][]{deblok08}. The effect of the constant density 
core may be remarkable for the evolution of  globular clusters in a
dwarf galaxy \citep[e.g.,][]{sanchezsalcedo06}, but not for the evolution 
of satellite halo in a Milky-Way sized halo.

\vspace{1cm}

\normalem
\begin{acknowledgements}
This work is funded by the National Natural Science Foundation
of China No. 10573028, the Key Project No. 10833005, the Group Innovation
Project No. 10821302, and by 973 program No. 2007CB815402.
XK is supported by the {\it One  Hundred Talents} project  of the
Chinese Academy of  Sciences and by the {\it foundation for the author 
of CAS excellent doctoral dissertation}. We thank the referee for 
constructive comments which significantly improve our paper.
\end{acknowledgements}

\label{lastpage}

\end{document}